\documentclass[
showpacs,superscriptaddress,twocolumn,aps, longbibliography]{revtex4-1} 

\usepackage[USenglish]{babel}

\usepackage{graphicx,xcolor}
\usepackage{amsmath, bbm}
\usepackage{mathtools}
\usepackage{natbib}
\usepackage{hyperref}
\usepackage{epstopdf}
\usepackage{color}
\usepackage{notes2bib}

\renewcommand{\imath}[0]{\mathrm{i}}

\newcommand{\myref}[1]{}

\begin{document}
\title{Partial cloaking of a gold particle by a single molecule}
 \author{Johannes Zirkelbach}%
\affiliation{Max Planck Institute for the Science of Light, D-91058 Erlangen, Germany}
 \author{Benjamin Gmeiner}%
\affiliation{Max Planck Institute for the Science of Light, D-91058 Erlangen, Germany}
\author{Jan Renger}%
\affiliation{Max Planck Institute for the Science of Light, D-91058 Erlangen, Germany}
\author{Pierre T\"urschmann}%
\affiliation{Max Planck Institute for the Science of Light, D-91058 Erlangen, Germany}
\author{Tobias Utikal}%
\affiliation{Max Planck Institute for the Science of Light, D-91058 Erlangen, Germany}
\author{Stephan G\"otzinger}%
\affiliation{Friedrich Alexander University Erlangen-Nuremberg, D-91058 Erlangen, Germany}
\affiliation{Max Planck Institute for the Science of Light, D-91058 Erlangen, Germany}
\affiliation{Graduate School in Advanced Optical Technologies (SAOT), Friedrich Alexander
University Erlangen-Nuremberg, D-91052 Erlangen, Germany}
\author{Vahid Sandoghdar}
\affiliation{Max Planck Institute for the Science of Light, D-91058 Erlangen, Germany}
\affiliation{Friedrich Alexander University Erlangen-Nuremberg, D-91058 Erlangen, Germany}

\date{\today}

\begin{abstract}
Extinction of light by material particles stems from losses incurred by absorption or scattering. The extinction cross section is usually treated as an additive quantity, leading to the exponential laws that govern the macroscopic attenuation of light. In this work, we demonstrate that the extinction cross section of a large gold nanoparticle can be substantially reduced, i.e., the particle becomes more transparent, if a single molecule is placed in its near field. This partial cloaking effect results from a coherent plasmonic interaction between the molecule and the nanoparticle, whereby each of them acts as a nano-antenna to modify the radiative properties of the other. 
\end{abstract} 

\maketitle

Macroscopic objects cast a shadow in a beam of light, and the shadow becomes darker if the medium is made optically thicker. This scenario also persists in the nanoscopic domain when the object is smaller than the wavelength of light. For instance, a gold nanoparticle (GNP) of diameter less than 100\,nm can extinguish more than half of the power from a green laser beam if placed in its focus \cite{Mojarad2009Metal, Celebrano2010Efficient}. According to the Beer-Lambert law, this shadow becomes exponentially darker as more particles are added \cite{Bohren1983Absorption}. However, it turns out that one can make a GNP \textit{transparent} to light by adding a single atom \cite{Wu2010Quantum, Ridolfo2010Quantum, Chen2013Coherent}. The underlying physics of this intriguing phenomenon lies in the interference between the fields scattered by the atom and the GNP in a near-field coupled configuration. Here, it is helpful to recall that the cross section of a two-level atom with transition at wavelength $\lambda$ can be as large as $\sigma_0=3\lambda^2/2\pi$ \cite{Zumofen2008Perfect}, which can be comparable to the extinction cross section and the physical size of a nanoparticle \cite{Bohren1983Absorption}. In other words, although both an atom and a GNP can individually extinguish a laser beam, their composite entity becomes transparent due to a coherent interference effect \cite{Chen2013Coherent}. 

Strong enhancement of light absorption or transmission by a plasmonic nanoparticle through coupling to a quantum emitter has recently been a subject of theoretical discussions \cite{Zhang2006Semiconductor, Artuso2008Optical, Wu2010Quantum, Ridolfo2010Quantum, Chen2013Coherent}. However, several factors make a laboratory demonstration of these effects challenging \cite{Hartsfield2015Single}. First, $\sigma_0$ is lowered and the homogeneous spectra are broadened by about five orders of magnitude for solid-state emitters at room temperature. Second, an emitter and a GNP would have to be placed at separations much smaller than a wavelength. Third, the orientation of the emitter dipole moment has to suit the geometrical features of the nanostructure. Despite decades of nanotechnology experience, these challenges are still not easy to tackle. In this Letter, we report on a successful realization of emitter-induced transparency using dye molecules at a temperature of T=1.5\,K, discuss how we overcome various experimental difficulties and validate our measurements using a theoretical model. 

\begin{figure}[t]
\centering 
 \includegraphics[width=1\columnwidth]{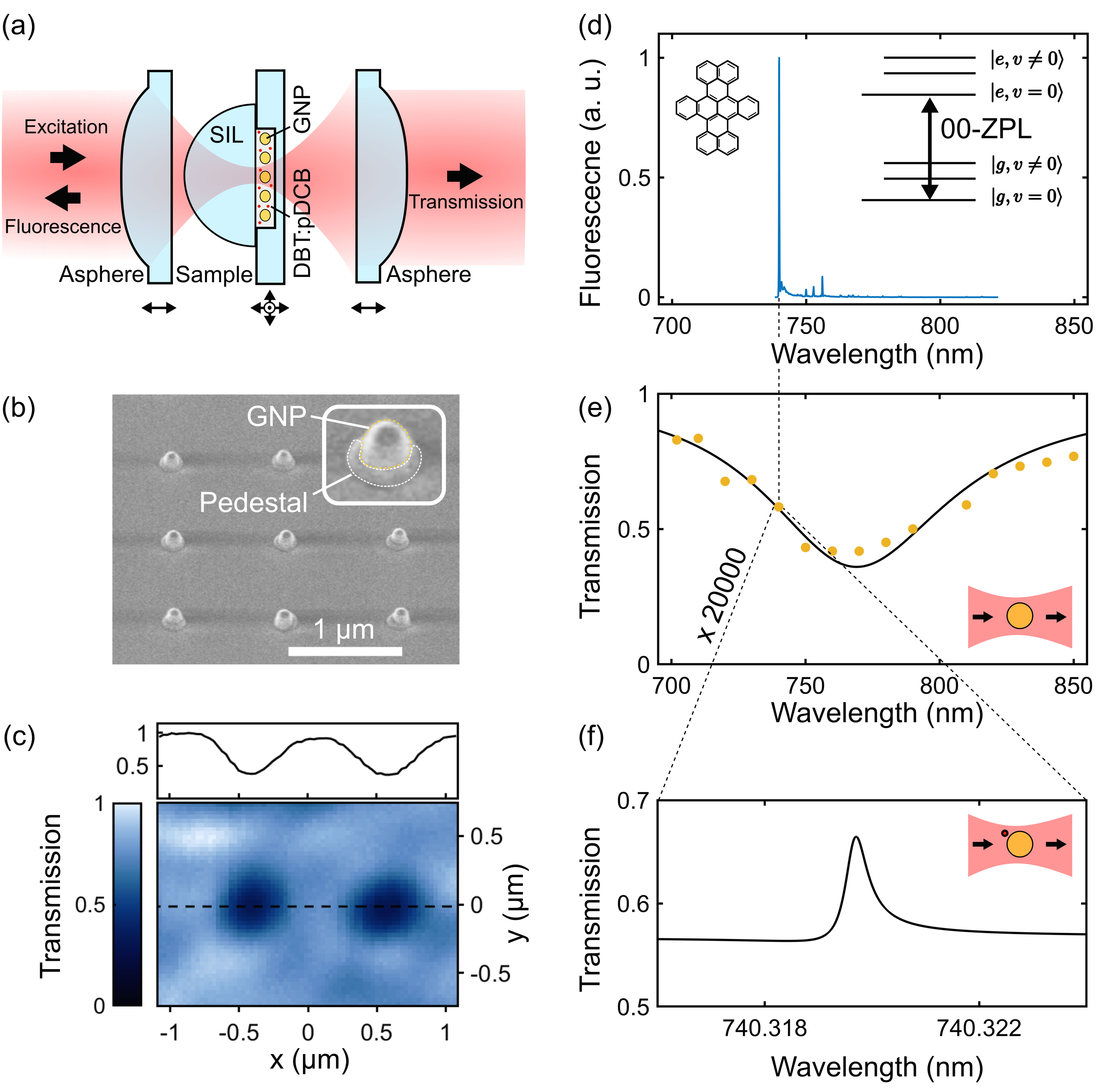}
 \caption{(a) Schematics of the core of the experimental arrangement. The sample consists of DBT molecules in a \textit{p}DCB crystal surrounding an array of gold nanoparticles prepared in a nanochannel (see text for details). SIL: solid-immersion lens, GNP: gold nanoparticle. The arrows at the bottom show the translational degrees of freedom for each component.
(b) Scanning electron microscope image of the GNP array. Inset: A zoom of a single GNP.
(c) Optical transmission image of two GNPs at $\lambda$=740\,nm. Upper panel shows a normalized cross section along the dashed line. (d) Fluorescence spectrum of DBT in \textit{p}DCB recorded upon excitation via transition from $\left|{g, v=0}\right>$ to $\left|{e, v\neq0}\right>$. The strong emission line at $\lambda=740.3$\,nm represents the 00ZPL. Inset: Molecular structure and Jablonski diagram of DBT. (e) Plasmon resonance of a GNP measured in transmission (dots) fitted by a Lorentzian profile (black curve). (f) Predicted reduction of the GNP extinction at the resonance of a single molecule that its coupled to it in the near field.}
\label{schematic}
\end{figure}

Figure\,\ref{schematic}(a) illustrates the schematics of the core of our experimental arrangement, where a laser beam is tightly focused onto a sample carrying GNPs and dibenzoterrylene (DBT) molecules. These are placed inside a channel of width 250\,$\mu$m and depth 245\,nm fabricated in a glass chip and covered by a ZrO$_{\rm 2}$ solid-immersion lens (SIL) with a diameter of 3\,mm. Figure\,\ref{schematic}(b) shows an electron microscope image of an array of GNPs with a diameter of about $130$\,nm, a height of about 100\,nm, and a spacing of 1\,$\mu$m. The GNPs are fabricated using electron beam lithography on evaporated gold films, followed by etching  and subsequent annealing, whereby the etch process is controlled to create glass pedestals of height 35\,nm under the GNPs (see Fig.\,\ref{schematic}(b) and the Supplementary Information, SI). A thin organic crystal of para-dichlorobenzene (\textit{p}DCB) lightly doped with DBT molecules is prepared to surround the GNP array by first introducing it into the channel in a molten state and next letting it solidify\,\cite{Gmeiner2016Spectroscopy} (see also SI). A 5\,nm layer of $\rm{Al_2O_3}$ is coated on the nanostructures by atomic layer deposition to provide a minimum separation between the DBT molecules and the gold to avoid strong quenching. The sample is then placed in a cryostat and cooled to $\rm T=1.5$\,K.

A beam from a Ti:Sapphire laser is coupled to the sample via an aspheric lens and the SIL, reaching a focus spot with a full width at half-maximum (FWHM) of 270\,nm assessed by mapping the fluorescence of a single molecule. A second aspheric lens is used to re-collimate the laser beam in transmission. Figure\,\ref{schematic}(c) displays a transmission image recorded by scanning the focus of the incident laser beam across two GNPs. A cut through the image (see upper panel) reveals an extinction dip of about 50\% from each GNP. In Fig.\,\ref{schematic}(e), we plot the extinction plasmon spectrum of a single GNP (see SI).

The inset in Fig.\,\ref{schematic}(d) displays the structure of DBT and its Jablonski diagram. DBT is a member of the polycyclic aromatic hydrocarbon (PAH) family and possess a strong zero-phonon line (00ZPL) between $\left|{g, v=0}\right>$ (ground electronic and vibrational state) and $\left|{e, v=0}\right>$ (ground vibrational and electronic excited state) when placed in an appropriate crystal. Figure\,\ref{schematic}(d) presents the emission spectrum of a single DBT molecule upon excitation to $\left|{e, v\neq0}\right>$ state, followed by a fast nonradiative decay to $\left|{e, v=0}\right>$ with a radiative lifetime of a few nanoseconds. The spectrum shows that a large fraction of the decay from $\left|{e, v=0}\right>$ takes place via the 00ZPL, leading to a branching ratio of about 44\% \cite{Verhart2016Spectroscopy}.

In our sample, DBT molecules are stochastically distributed in the \textit{p}DCB matrix, but we can identify and interrogate each molecule individually with very high spatial and spectral resolution. Here, we first scan the wavelength of the narrow-band laser across the inhomogeneous band of DBT:\textit{p}DCB around $\lambda$=740\,nm \cite{Gmeiner2016Spectroscopy, Verhart2016Spectroscopy}. The exquisitely narrow 00ZPL resonances associated with the molecules do not overlap so that each can be selectively addressed by tuning the laser frequency. The spectral selection of a single molecule also allows us to image it and, thus, determine the center of its point-spread function (PSF) beyond the diffraction limit. 

Our goal in this Letter is to show that a single molecule can counteract the extinction effect of a single GNP. Our strategy is first, to locate a DBT molecule close to a GNP and examine its near-field coupling via incoherent fluorescence measurements. We then explore the coherent effect of the composite system of molecule-GNP. As depicted in Fig.\,\ref{schematic}(f), we expect the resonant transmission signal to experience a substantial increase.

\begin{figure}[t]
\centering 
 \includegraphics[width=1\columnwidth]{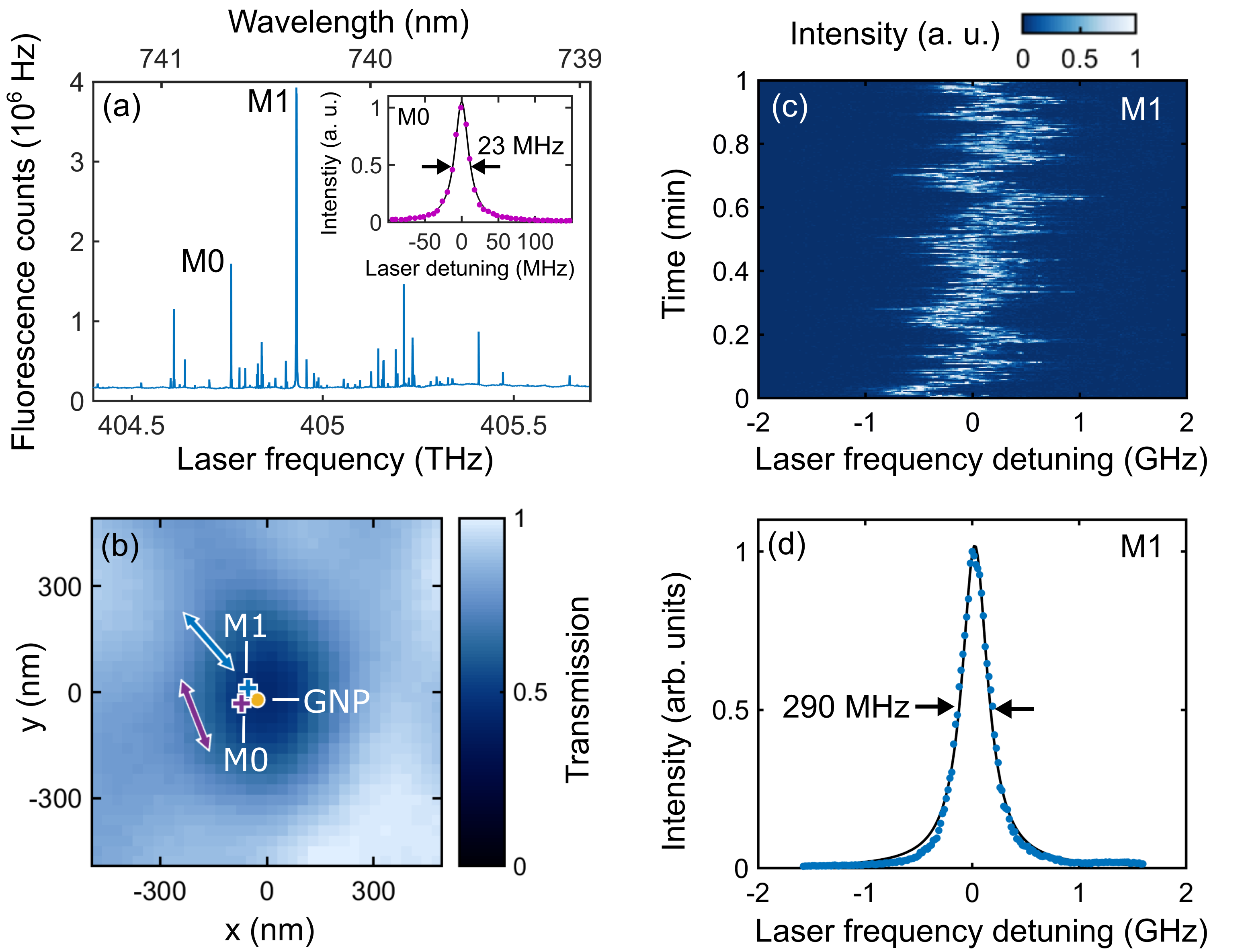}
 \caption{(a) Fluorescence of DBT molecules located within a focal spot of one GNP as a function of the excitation laser frequency. The applied laser power of 79\,nW corresponds to an excitation intensity above saturation. Inset: A zoom onto the spectrum of M0 recorded at low excitation power.
(b) An extinction image of a single GNP recorded in transmission overlaid with the locations of the GNP obtained from its fluorescence image (yellow), M0 (magenta) and M1 (blue). The arrows depict the in-plane orientations of the molecular dipole moments. (c) 240 spectra recorded at 20\,GHz/s over one minute well below saturation (excitation power 0.04 nW). 
(d) Sum of the aligned spectra shown in (c) (symbols) and a Lorentzian fit (solid curve).}
\label{spectra}
\end{figure}

To identify molecules that are located in the near field of a GNP, we first centered the focus of the laser beam on the GNP and scanned the laser frequency. Figure\,\ref{spectra}(a) presents an example of the ZPLs obtained at this position by recording the red-shifted fluorescence as the laser frequency was scanned through the inhomogeneous band of about 1\,THz. The differences in the observed signal stem mostly from variations in the positions of the molecules within the laser intensity profile. The inset in Fig.\,\ref{spectra}(a) displays a zoom into one of the strongest ZPL resonances (marked as M0) with a linewidth of 23\,MHz measured at low excitation power. This spectrum is consistent with those reported in bulk crystals \cite{Verhart2016Spectroscopy} and represents a typical spectral response of the molecules that are not coupled to GNPs in our current sample. 

We used a conventional PSF localization procedure to determine the lateral positions of all the molecules associated with the different resonances in Fig.\,\ref{spectra}(a) at a precision of about 10\,nm. In addition, we applied this method to the fluorescence image of the GNP \cite{Fang2012Plasmon,Cheng2016Luminescence}.  The yellow circle and the magenta cross in Fig.\,\ref{spectra}(b) mark the lateral positions of the GNP and M0, respectively, but it should be borne in mind that the results can contain systematic errors due to the redirection of the molecular emission by the GNP \cite{Su2016Super, Fu2017Super, Blanquer2020Relocating}. 

Molecules that experience a significant degree of plasmonic enhancement are expected to display shorter fluorescence lifetimes, broader 00ZPLs and higher emission rates upon saturation. In what follows, we investigate one such molecule, which we label M1, in great detail. The blue cross in Fig.\,\ref{spectra}(b) marks the apparent center of M1 overlaid on the extinction image of the GNP. These measurements indicate that the GNP, M0 and M1 are laterally very close to each other. The channel depth of 245\,nm limits the axial distance variations and greatly simplifies the search for GNP-coupled molecules. We also performed polarization studies to gain insight into the orientation of the molecular transition dipole moment. We found the in-plane dipole moment of M1 to point nearly radially towards the GNP (considered to be spherical) whereas the dipole moment of M0 was tangential to the GNP (see Fig.\,\ref{spectra}(b)).

Repeated fluorescence excitation spectra in Fig.\,\ref{spectra}(c) reveal that M1 experienced spectral instabilities over a time scale of seconds. Indeed, spectral diffusion poses one of the great challenges in performing high-resolution coherent studies in the near field of plasmonic structures because the crystallinity of solid-state matrices, be it semiconductor, inorganic or organic is compromised at an interface between different materials and geometries. To get around the spectral jitter, we scanned the laser frequency at a fast rate of 20\,GHz/s. The blue symbols in Fig.\,\ref{spectra}(d) show the average of 240 such scans obtained after aligning their midpoints determined as the center of the measured FWHMs in each scan. A good Lorentzian fit (black curve) indicates that our procedure successfully accounts for the slow spectral diffusion. We also applied a similar analysis to account for the residual spectral diffusion that occurred in some measurements on M0.

\begin{figure}[t]
\centering 
 \includegraphics[width=1\columnwidth]{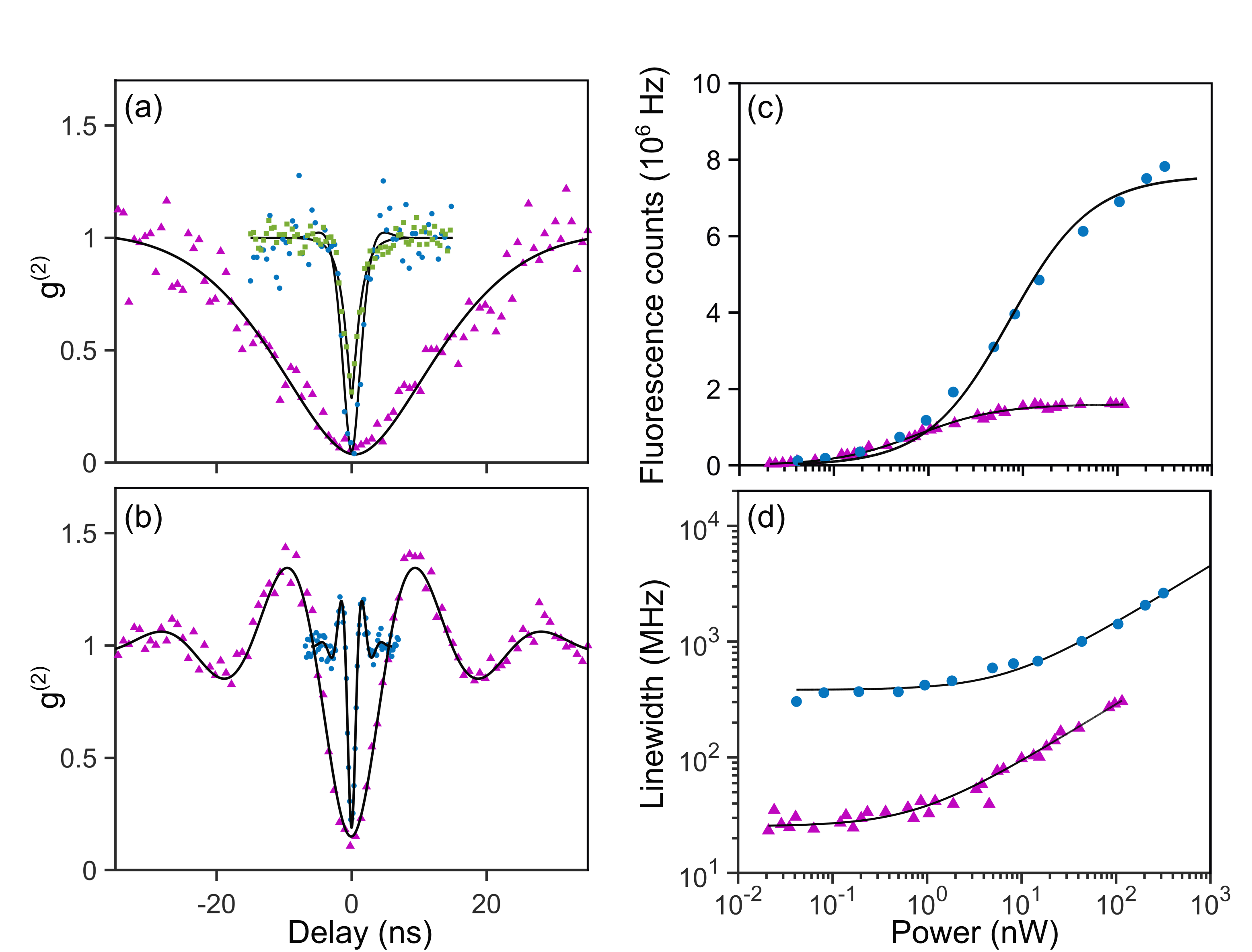}
 \caption{(a) Second-order autocorrelation function, $g^{(2)}(\tau)$ for M0 excited via 00ZPL (magenta), M1 excited via 00ZPL (blue) and M1 excited via a higher vibrational level $\left|{e, v\neq0}\right>$ (green). (b) Fluorescence signal versus excitation power for M0 (magenta; max value: ($1.6\pm0.1) \times 10^6$ counts per second) and M1 (blue; max value: ($7.7\pm0.1) \times 10^6$ counts per second) under 00ZPL excitation. (c) 00ZPL linewidth (FWHM) versus excitation power extracted from the spectra used in (b). (d) Same as in (a) but excited above saturation. }
\label{g2-sat}
\end{figure}

The extracted FWHM of 290\,MHz is much broader than that of M0 and provides a first indication for plasmonic coupling. To examine this further, we measured fluorescence intensity correlations in a Hanbury-Brown and Twiss arrangement. The magenta triangles in Fig.\,\ref{g2-sat}(a) display $g^{(2)}(\tau)$ for M0 excited via its 00ZPL, while the blue and green dots show the outcome for M1 excited via its 00ZPL and a higher vibrational level of the electronic excited state $\left|{e, v\neq0}\right>$, respectively. Pronounced antibunching effects at zero delay time assure us that the signals stem from single molecules. A fit to the measured data lets us extract an excited state lifetime of $T_1 = 8.1 \pm 0.4\,$\,ns for M0 and $T_1 = 1.4 \pm 0.1\,$\,ns for M1 (see SI). The resulting modest six-fold lifetime shortening points to a plasmonic Purcell effect \cite{Kuhn2006Enhancement, Sandoghdar2013Antennas}. 

A shorter excited-state lifetime could result from both the enhancement of the radiative ($\Gamma_{\rm r}$) and nonradiative ($\Gamma_{\rm nr}$) rates. To inquire about the relative weights of these effects, we excited M0 and M1 via their 00ZPLs at different incident powers. The fluorescence signals presented in Fig.\,\ref{g2-sat}(b) show that at saturation, the power radiated by M1 is about five times larger than that of M0 if we assume similar collection efficiencies\,\bibnote{This is a reasonable assumption because the two molecules are very close to each other and the numerical aperture of our collection optics is very high.}. This confirms a substantial Purcell enhancement of the radiative decay. Figure\,\ref{g2-sat}(c) also presents the evolution of the molecular linewidths as a function of the laser power. 

Our findings verify that a GNP acts as a plasmonic nanoantenna to enhance the radiative properties of M1. However, as in the case of the great majority of previous reports on plasmonic antennas \cite{Kuhn2006Enhancement, Sandoghdar2013Antennas, Matsuzaki2017Strong, Koenderink2017Single}, the above-mentioned studies were solely based on the behavior of the excited-state population observed via the red-shifted fluorescence. Resonant scattering, however, depends sensitively on the degree of coherence in the molecular dipole moment. Indeed, the measured fluorescence lifetime of 1.4\,ns lets us deduce a homogeneous linewidth ($\Gamma_1/2\pi$) of 114 $\pm$ 8 \,MHz for the 00ZPL of M1, which is notably less than the directly measured FWHM of 290\,MHz. Thus, we expect a contribution from pure dephasing. 

To extract additional information about the fast dynamics that might contribute to dephasing, we analyzed $g^{(2)}(\tau)$ as a function of excitation power. Figure\,\ref{g2-sat}(d) plots two examples of such measurements for M0 (magenta) and M1 (blue) under strong excitation. A simultaneous fit of the data let us extract a pure dephasing rate of $\Gamma^{\star}/2\pi=87 \pm 35$\,MHz for M1, corresponding to $T_2^{\star}\sim1.8$\,ns (see SI for a detailed discussion). This implies $\rm FWHM=(\Gamma_1  + 2\Gamma_2 ^{\star})/2\pi=(114 + 2\times87)\rm MHz=288$\,MHz, which is in good agreement with the directly measured value of 290\,MHz. 

\begin{figure}[t]
\centering 
 \includegraphics[width=1\columnwidth]{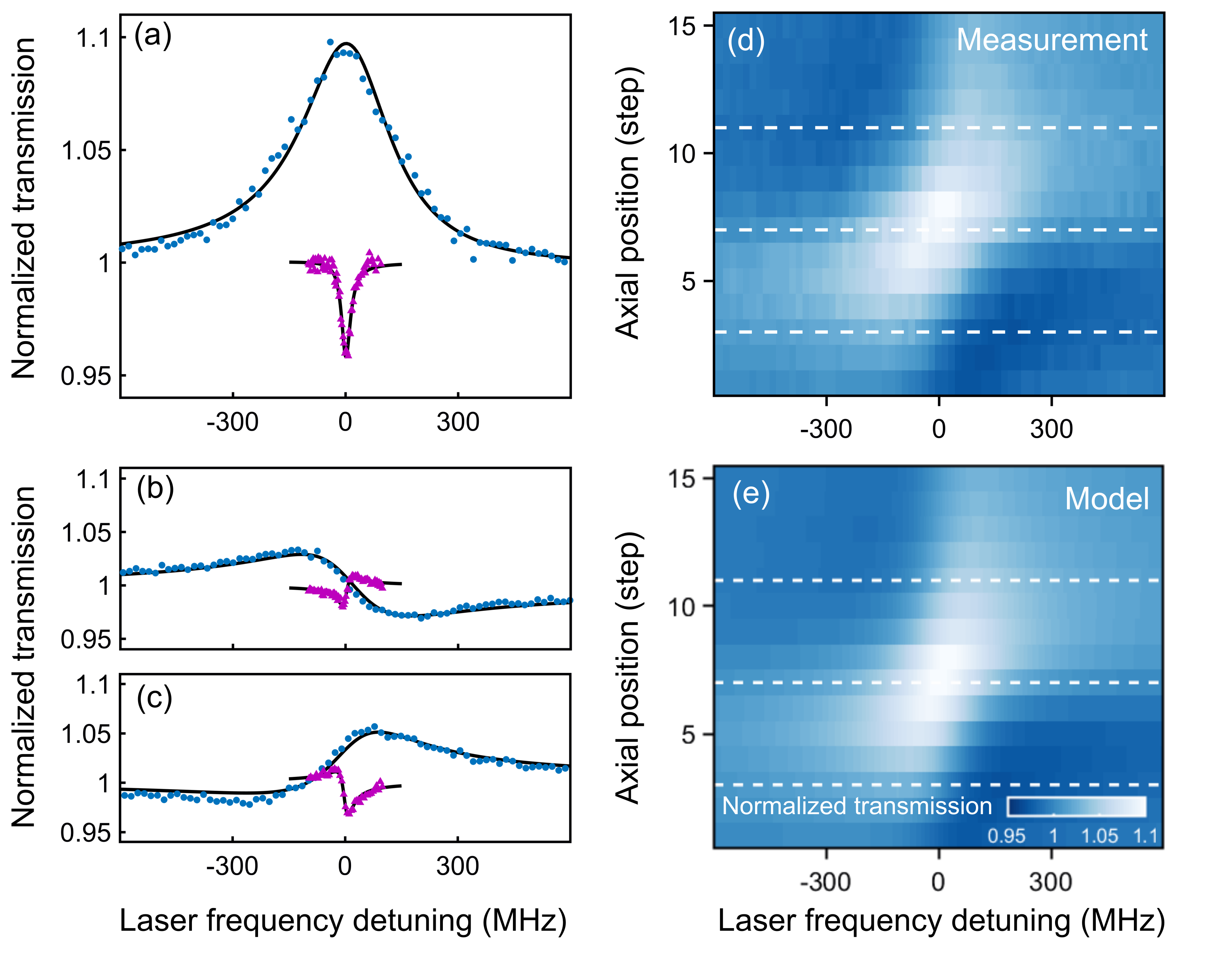}
 \caption{(a) Transmission spectra recorded about the 00ZPL resonances of M1 (blue) and M0 (magenta) placed in the focus of the excitation beam, corresponding to the middle dashed cut in (d). (b,c) Same as in (a) but for molecules placed on two opposite sides of the focus, as indicated by the upper and lower dashed cuts in (d). (d) Series of transmission scans through the 00ZPL resonance of M1 at 15 different axial positions of the sample over an estimated total distance of 3.5\,$\mu$m (see SI). (e) Calculated transmission spectra to fit the data in (d). The solid curves in (a,b,c) correspond to the three cuts marked by the dashed lines. See SI for details.} 
\label{focus-z}
\end{figure}

Having established a good understanding of the degree of coherence in our molecular system, we now present our results on resonant extinction spectra. The magenta data points in Fig.\,\ref{focus-z}(a) display the transmitted power of the incident laser beam as its frequency was scanned through the 00ZPL of M0. The observed extinction dip of 4\% is in the range of previous measurements on single PAHs \cite{Wrigge2008Efficient}\bibnote{We observed extinction dips as large as 10\% for other molecules in our sample. The lower extinction effect is because the GNP in this study did not lie on the SIL axis.}. However, the blue symbols in Fig.\,\ref{focus-z}(a) show that in the case of M1, in addition to a larger linewidth, the transmitted power is \textit{increased} by 10\%. In other words, by adding a single molecule, we have indeed turned a gold nanoparticle more transparent. 

Because extinction is intrinsically an interferometric phenomenon \cite{Zumofen2008Perfect}, the laser intensity observed in the far field depends on the relative phase and amplitude of the excitation and scattered fields at the detector. Thus, considering the characteristic dependence of the Gouy phase around the focal plane, one expects a clear change in the resonance profile upon axial scan of the sample \cite{HwangOptComm2007}. The Fano-type spectra in Fig.\,\ref{focus-z}(b,c) show this effect for both M0 (magenta) and M1 (blue) at two positions of the sample-SIL assembly across the focal plane of the aspherical lens (see Fig.\,\ref{schematic}(a)). In Fig.\,\ref{focus-z}(d) we plot the extinction spectra of M1 for an extended axial sweep. A more detailed discussion and the equivalent data for M0 can be found in the SI.

The quantitative details of the plasmonic coupling of M1 crucially depend on the geometrical features of the GNP as well as the exact position and orientation of the molecule. Considering that we do not have access to these parameters, we cannot use rigorous numerical simulations to fit our experimental data. However, the underlying physics can be captured by employing a model based on driven coupled oscillators \cite{Garrido2002Classical, Gallinet2018Model, Ruesink2018Controlling} (see SI for details). Figure \,\ref{focus-z}(e) presents the outcome of such calculations fitted to the corresponding measurements shown in Fig.\,\ref{focus-z}(d). The solid curves in Fig.\,\ref{focus-z}(a--c) through the measured data of M1 represent the cuts marked by the dashed lines in Fig.\,\ref{focus-z}(e). The consistent agreement between the experimental data and the theoretical model provides assurance that the resonance profiles follow the phase behavior expected from the coherent interaction between the laser beam, GNP and M1. We note that our model also predicts a ``Lamb shift" of 12\,MHz in the resonance of M1 induced by coupling to the GNP. While this small frequency shift is accessible to our high-resolution spectroscopic studies, we did not verify it because we could not examine M1 in the absence of GNP as is done in scanning-probe arrangements \cite{Kuhn2006Enhancement,Hartsfield2015Single,Matsuzaki2017Strong, Gross2018Near,Park2019Tip}.

Over the past fifteen years, plasmonic platforms have been employed in a wide range of studies for modifying the optical properties of quantum emitters \cite{Koenderink2017Single}. The main thrust of these works has been in the \textit{incoherent} enhancement of excitation or fluorescence rates, where the metallic nanostructure at hand acts as an optical antenna for ameliorating the efficiency of interaction between the emitter and propagating photonic modes \cite{Sandoghdar2013Antennas}. Recently, there have also been the first reports of coherent plasmonic interactions, including emitters placed in extremely small gaps of plasmonic antennas to reach the strong coupling regime of Cavity Quantum Electrodynamic (CQED)\cite{Hartsfield2015Single,Chikkaraddy2016Nature, Gross2018Near, Park2019Tip, Pelton2019Strong}. The degree of coherence achieved in these experiments, however, has been very limited because they were performed at room temperature. By operating at cryogenic conditions, we have increased the coherence time at play by about five orders of magnitude \bibnote{Typical linewidths of a solid-state emitter at our wavelength are about 10\,THz at room temperature. We reach linewidths in the order of 100\,MHz.}, essentially reaching the natural linewidth limit of the emitter. 

The coherent interference of the fields scattered by a single organic molecule and a gold nanoparticle via near-field plasmonic interactions is analogous to the effect of a quantum emitter on the transmission of a microresonator in the weak coupling regime of CQED \cite{Wang2017Coherent}, which has also been termed ``dipole-induced transparency" \cite{Waks2006Dipole}. Plasmonic antennas, however, offer several advantages over conventional microresonators, including their nanoscopic compactness and broad bandwidth. These features provide important opportunities for the realization of subwavelength building blocks of quantum photonic circuits for key operations such as switching and phase modulation. In future, stronger near-field couplings through more advanced antenna designs \cite{Sandoghdar2013Antennas, Chikkaraddy2016Nature, Matsuzaki2017Strong} will enhance the system performance well beyond the first demonstration presented in this Letter.

\textbf{Acknowledgments}
We thank Jaesuk Hwang and Andreas Maser for their very early experimental efforts on the project. This work was supported by an Alexander von Humboldt professorship and the Max Planck Society. 

%

\end{document}